\documentstyle[11pt,aaspp4]{article}

\newcommand{\mr}{MR~2251--178}
\newcommand{\pasa}{{\rm PASA}}
\newcommand{\etal}{{\it et al.\/}}
\newcommand{\kms}{\hbox{km~s$^{-1}$}}
\newcommand{\ha}{\hbox{H$\alpha$}}
\newcommand{\hi}{\hbox{{\ion{H}{1}}}}
\newcommand{\hii}{\hbox{{\ion{H}{2}}}}
\newcommand{\oiii}{\hbox{[{\ion{O}{3}}]}}

\slugcomment{Accepted by {\it The Astrophysical Journal Letters\/}.}

\lefthead{Shopbell \etal} \righthead{Ionized Nebula around MR2251-178}

\begin{document}

\title{The Very Extended Ionized Nebula around the Quasar MR2251-178}

\author{P. L. Shopbell, S. Veilleux\altaffilmark{1}}
\affil{Department of Astronomy, University of Maryland, College Park,
MD 20742; pls,veilleux@astro.umd.edu}
\and
\author{J. Bland-Hawthorn}
\affil{Anglo-Australian Observatory, P.O. Box 296, Epping, NSW 2121,
  Australia; jbh@aaoepp2.aao.gov.au}

\altaffiltext{1}{Cottrell Scholar of the Research Corporation}

\begin{abstract}
We report the results of deep \ha\ imaging of the ionized gas
surrounding the low-redshift ($z=0.0638$) quasar \mr\ using the TAURUS
Tunable Filter (TTF) on the Anglo-Australian Telescope.  Our
observations reach a 2-$\sigma$ detection level of $\sim$ 5 $\times$
10$^{-18}$ erg s$^{-1}$ cm$^{-2}$ arcsec$^{-2}$, more than an order of
magnitude deeper than conventional narrowband images previously
published on this object.  Our data reveal a spiral complex that
extends more or less symmetrically over $\sim200$~kpc, making it the
largest known quasar nebula.  The total mass of ionized gas is
$6\times10^{10}$~M$_{\sun}$ (upper limit), a large fraction of which
is in a very faint, diffuse component.  The large and symmetric extent
of the gaseous envelope favors a model in which the filamentary and
diffuse emission arises from a large cloud complex, photoionized by
the bright quasar.  A crude kinematic analysis reveals relatively
smooth rotation, suggesting that the envelope did not originate with a
cooling flow, a past merger event, or an interaction with the nearby
galaxy G1.
\end{abstract}


\section{Introduction}
\label{sec:introduction}
In the ongoing quest to better understand the luminous quasar
population, one of the most fertile areas of study has been the
investigation of their galaxy environments.  The host galaxy not only
must provide fueling for the central engine, but should also display
the effects of the strong nuclear ionizing emission and the impact of
dynamic activity, such as jets and winds.

Broadband observations of quasar environments suggest a surprising
variety of host galaxy morphologies, including spirals and
ellipticals, nearby companions, and tidal interactions.  While early
ground-based work indicated a large percentage of spirals amongst
low-redshift quasar hosts ($\gtrsim 40$\%; \cite{HCC84}) and no
concrete evidence for ellipticals, subsequent HST observations find an
elliptical fraction of greater than half (\cite{BKSS97}).  More recent
HST work suggests in fact that nearly all radio-loud and radio-quiet
quasars reside in massive ellipticals (\cite{MDKBOH99}; see also
\cite{DBBBCDMMSP95}).  A substantial fraction of quasar host galaxies
exhibit tidal tails and streamers indicative of interactions (e.g.,
\cite{HC83}; \cite{HCCDG84}; \cite{SM87}; \cite{BKSS97};
\cite{HCMDS99}).  Evidence for extended gas from interactions has also
been found at \hi\ 21 cm (e.g., \cite{LH99}). The presence of
companion galaxies is common; the QSO/galaxy correlation function is
evidently several times that for galaxies alone (\cite{FBKS96}).

Given the strong line emission in the nuclei of active galaxies and
quasars, as well as the prevalence of ionized gas in interacting
systems, narrowband observations of quasar environments have proved
interesting as well.  The imaging survey of Stockton \& MacKenty
(1987), the largest to date, found highly-structured \oiii\ emission
in a quarter of 47 luminous QSOs, with typical extents of a few tens
of kpc.  Similar spectroscopic observations have detected extended
regions of line emission in half of the objects observed
(\cite{BPO85}).  The line ratios usually suggest photoionization by
the nuclear power-law spectrum (e.g., \cite{BPO85}; \cite{BDBP94}),
although stellar absorption lines from (presumably in-situ) stars are
sometimes seen (e.g., \cite{MTS96}).  Kinematic studies of the ionized
gas component are scarce, but the gas motions generally appear to be
complex and chaotic (e.g., \cite{DPPB94}). The size and luminosity of
the ionized component appear to be correlated with both the
narrow-line nuclear luminosity and the radio power of the quasar
(\cite{BPO85}; \cite{SM87}; \cite{VW90}).

The quasar \mr\ is one of a few radio-quiet quasars which exhibit an
extended gaseous envelope (e.g., \cite{BBDT83}).  This quasar was
first discovered as a strong and variable X-ray source by the {\it
Ariel V\/} satellite (\cite{CRMPEWGPMMSPT78}).  Subsequent
observations identified the object as a quasar at a redshift of
$0.0638\pm 0.0015$ (\cite{RCDDJDMH78}; \cite{CMR78}), residing in the
outskirts of a small cluster (\cite{P80}).  In this {\em Letter}, we
present deep \ha\ observations of \mr\ obtained with the TAURUS
Tunable Filter (TTF), a new, etalon-based instrument which has been
optimized for the detection of faint, extended emission-line gas.
These new data allow us to better constrain the extent, velocity
field, and origin of the ionized nebula around \mr.

\section{Observations and Reductions}
\label{sec:obsred}

\mr\ was observed on August 30, 1998 using the TTF at the f/8
Cassegrain focus of the 3.9-meter AAT. The TTF instrument consists of
a pair of modified high-finesse ($N \sim 40$) Queensgate etalons (blue
and red) which can be tuned to provide narrowband imaging anywhere
within the wavelength range 400 to 960~nm, through an arbitrary
bandpass, with resolving powers of 100 to 1000.\footnote{Consult
Bland-Hawthorn \& Jones (1998) and the TTF web page ({\tt
http://www.aao.gov.au/local/www/jbh/ttf/}) for further details.} The
observations of \mr\ were made with the red side of the TTF, using a
mediumband ($\Delta\lambda = 26.0$~nm) blocking filter centered at
707~nm, tilted by 16\arcdeg.  Two 600-second exposures were obtained
at each of two etalon spacings.  The exposures were dithered amongst
pointings on a 15\arcsec\ grid.  The average atmospheric seeing of
1.3\arcsec\ was oversampled by the 0.37\arcsec\ pixels.  The night was
photometric.  \mr\ was also observed on September 3, 1998 in direct
imaging mode, using a standard I-band filter.

Fits to a number of emission lines from observations of a calibration 
lamp (CuAr) were used to determine the relationship between 
wavelength, spatial position, and etalon gap spacing.  The free 
spectral range of the etalon, i.e., the distance between orders, was 
found to be 265.8~\AA, well-matched to the bandwidth of the blocking 
filter.  The system was used in the 26th order of interference, with a 
9~micron etalon gap, at an effective finesse of 39.7.  This translates 
to a spectral resolution of 6.8~\AA, or $R \sim 1040$, with an 
effective bandpass of $\sim12$~\AA. The pair of etalon spacings 
produced imagery with central wavelengths on the optical axis of 
6983~\AA\ and 6986~\AA, corresponding to redshifted velocities of 60 
and 195~\kms\ relative to the quasar.  The field is essentially 
monochromatic: pixels $\sim 1$\arcmin\ from the optical axis have a 
central wavelength only 1.3~\AA\ ($60$~\kms) bluer than pixels 
on-axis.  The optical axis is located approximately 35\arcsec\ 
northwest of the quasar.

The data frames were bias-subtracted and flatfielded in the typical
manner.  An azimuthally-symmetric sky frame was produced from a mean
radial sky spectrum for each image and subsequently subtracted.
Images at each of the two etalon spacings were then aligned and
combined using 9 stellar objects in the field.  An image mask was used
to simultaneously remove three ghost reflections from each image.  The
data were flux calibrated using observations of the standard star
EG~21 (\cite{SB83}).

The I-band observations were reduced following standard CCD
procedures.  The summed continuum image was matched to the narrowband
imagery using a number of stars and subsequently subtracted.  A few
remaining stellar residuals and cosmetic defects were identified and
repaired by hand.

\section{Discussion}
\label{sec:discussion}
Figure~\ref{fig:ttf_ha} presents our final \ha\ images of \mr\ at each
of the two etalon spacings (panels $a$ and $b$), as well as the I-band
continuum image (panel $c$) and a combined \ha\ image (panel $d$).
The 2-$\sigma$ detection limit of the H$\alpha$ images is $\sim$ 5
$\times$ 10$^{-18}$ erg s$^{-1}$ cm$^{-2}$ arcsec$^{-2}$, more than an
order of magnitude fainter than previously published data on this
object.  In the discussion that follows, we assume $H_0 =
50$~km~s$^{-1}$~Mpc$^{-1}$ and a corresponding image scale of
1.9~kpc~arcsec$^{-1}$.

\subsection{Morphology}
\label{sec:morphology}
Published optical imagery and spectroscopy have detected two ionized
gas components around the quasar \mr: an elongated, highly-ionized
circumnuclear component of diameter $\sim27$~kpc, and an extended
envelope of faint \ha- and \oiii-emitting filaments out to a radius of
$\sim 110$~kpc (\cite{BBDT83}; \cite{HNJ84}; \cite{APM84};
\cite{MCGPA90}). Our observations confirm these findings and reveal a
number of new features.

The circumnuclear component of ionized gas is evidenced by strong 
\ha\ emission directly surrounding the quasar, extending out to a 
radius of $\sim 20$~kpc.  This emission is slightly elongated in the 
east-west direction, i.e., along the axis of the quasar's radio jet 
($PA \sim 102$\arcdeg; \cite{MCGPA90}), and has been identified as an 
``extended emission-line region'' (EELR) of the quasar host galaxy 
(\cite{MV92}).  As others have reported on characteristics of the EELR 
in some detail, we will say nothing more of it herein.

The extended ionized gas component is observed as a much larger
network of diffuse and filamentary emission exterior to the EELR.
Above a flux level of $1.8 \times
10^{-17}$~erg~s$^{-1}$~cm$^{-2}$~arcsec$^{-2}$, diffuse \ha\ emission
is visible surrounding the galaxy to a radius of at least 50~kpc.
Knots and filaments of ionized gas are visible to radii up to 120~kpc
towards the northeast, southeast, and southwest.  The filaments
display a spiral morphology and exhibit a remarkable azimuthal
symmetry that has not been previously seen. The total extent of this
complex ($\sim 200$~kpc) makes it the largest known around a quasar.
We derive a total flux from the extended ionized gas component of $1.4
\times 10^{-13}$~erg~s$^{-1}$~cm$^{-2}$, or a luminosity of $2.5
\times 10^{42}$~erg~s$^{-1}$ at the redshift distance of the quasar
(384~Mpc).  This value, as well as the corresponding upper limit on
the gas mass of $6\times10^{10}$~M$_{\sun}$, is comparable or slightly
greater than previously published results (e.g., \cite{NHJC86};
\cite{MCGPA90}).

A few of the individual knots and filaments have been previously
noted; these are indicated in Figure~\ref{fig:ttf_ha}$d$, using the
denotations from Macchetto \etal\ (1990).  Several additional knots
also noted by these authors and others (e.g., \cite{APM84}) are
conspicuous in their absence from our imagery.  Most notable among
these are the ``E'' knots interior to the eastern filament in
Figure~\ref{fig:ttf_ha}.  As will be noted in the following section,
this absence is probably due to the large range of velocity spanned by
the filament emission, relative to our narrow bandpasses.

\subsection{Kinematics}
\label{sec:kinematics}
The spectroscopy which initially discovered the extended ionized
component of \mr\ also provided the first kinematical information on
the nebula (\cite{BBDT83}). Those observations revealed a normal
rotation curve along each of three position angles, with peak
velocities of $\sim 150$~\kms. Surprisingly, the extended nebula
appears to be rotating in an opposite sense to the inner (EELR)
regions of the galaxy (e.g., \cite{NHJC86}). The only published
two-dimensional kinematics of the system are narrow-field \ha\
Fabry-Perot observations of the EELR (\cite{MV92}).

Comparison of our \ha\ imagery at each of the two distinct etalon 
positions reveals a broad velocity gradient across the quasar nebula.  
The maximum gradient appears to be along a position angle of $\sim 
40$\arcdeg, with the eastern filaments generally less redshifted 
than the western filaments.

The eastern filaments are clearly detected in both the ``blue'' and
``red'' images.  Since an upper limit of $\sim200$~\kms\ has been
measured for their line widths (\cite{BBDT83}), the central velocity
of these filaments must lie within the overlap region of our two
velocity regimes.  Accounting for its distance from the optical axis,
the ionized gas east of the quasar must be near the systemic velocity
of the quasar.

The western filaments are detected only in the ``red'' image and 
appear to be redshifted from the systemic velocity by a value of $\sim 
200-300$~\kms.  This correlates well with the imaging observations of 
Macchetto \etal\ (1990), whose broad ($\Delta\lambda = 64$~\AA) \ha\ 
filter barely detected the western filaments out to a velocity of 
$\sim200$~\kms\ relative to the quasar.  Furthermore, these authors 
more readily detected the western filaments in their \oiii\ imagery, 
which encompassed a redder range of velocities.

We therefore confirm that the velocity structure of the extended
ionized gas component does not appear to follow the rotation curve of
the quasar (\cite{NHJC86}; \cite{MV92}).  The broad velocity range of
our images and the limitation of two distinct velocity samples
restricts us from analyzing more precise velocity variations, e.g.,
amongst individual knots and filaments.

\subsection{The Origin of the \hii\ Envelope}
\label{sec:origin}
Several scenarios have been put forth for the origin of the ionized
gas envelope in \mr, including tidal debris from an interaction with
galaxy G1 some $\sim2 \times 10^8$~years ago, similar debris from a
merging event, clouds expelled from the quasar and/or host galaxy, a
cooling flow, or the ionized portion of a large \hi\ envelope that is
gravitationally bound to the quasar (e.g., \cite{NHJC86};
\cite{MCGPA90}).  Our observations reveal a tremendous spatial extent
for the emission in this system.  We also observe a relatively high
degree of symmetry in the envelope, including at least two ``arms'' of
ionized gas and a pervasive diffuse component.  This symmetry,
together with the organized large-scale kinematics, casts doubt upon
the interaction model for creation of the ionized gas.  Furthermore,
the galaxy G1, the usual culprit quoted as an interacting companion to
\mr, has a velocity redshifted by 1246~\kms\ relative to the quasar
(\cite{BBDT83}), yet its spatial position corresponds to that of
relatively blueshifted filaments. Published spectroscopy shows the
emission lines from these filaments to be symmetric and significantly
narrower than this (e.g., \cite{BBDT83}), although we cannot rule out
the presence of a very faint, high-velocity streamer extending out
toward G1.

Nevertheless, the coarse kinematic structure of the extended ionized
gas indicates that it represents a distinct component from the inner
EELR region, since the latter appears to rotate in an opposite sense.
This implies that the envelope most probably did not originate within
the host galaxy of the quasar.  A retrograde merging event, in which a
small, gas-rich galaxy has been subsumed by the quasar, remains a
possibility, however again the relatively well-ordered large-scale
kinematics and azimuthal symmetry of the ionized envelope are
problematic.

The cooling flow hypothesis appears equally unlikely. As pointed out
by Macchetto \etal\ (1990), the large size and ordered kinematics of
the envelope, and the off-centered position and peculiar systemic
velocity of the quasar with respect to the underlying cluster, all
argue against this scenario.  By elimination, our observations
therefore favor a model in which the extended ionized envelope resides
within a large complex of \hi\ gas centered about the quasar.  If this
is correct, a deep search for 21~cm \hi\ line emission should reveal
the massive neutral envelope and constrain its origin (e.g., captured
intergalactic \hi\ clouds, remnant accreting gas from galaxy
formation, etc.).

\subsection{The Ionization of the \hii\ Envelope}
\label{sec:ionization}
The high excitation level of the brighter knots around \mr\ (e.g.,
\cite{BBDT83}; \cite{NHJC86}; \cite{MCGPA90}) requires an energetic
source of ionization.  In-situ ionization by a faint hot stellar
component is unlikely to be dominant for a number of reasons.  Deep
broadband imagery of \mr\ reveals a morphology very unlike that of our
narrowband observations.  The R-band image of Hutchings \etal\ (1999)
exhibits a faint component extended along the N--S direction, markedly
different from the symmetric spiral pattern that we observe in \ha.
Moreover, Hutchings \etal\ (1999) do not detect an excess of continuum
emission at the locations of any of the brighter \ha\ knots in the
envelope.  Perhaps even more damaging to the stellar ionization
scenario is our detection of a large amount of diffuse ionized gas
outside of the brighter knots.  We therefore favor an external source
of ionization.

The presence of a significant amount of ionized gas at large angles 
relative to the jet axis (PA $\sim 102$\arcdeg; \cite{MCGPA90}) seems 
to rule out ionization mechanisms directly associated with the quasar 
jet (e.g., shocks).  We are therefore left with the possibility that 
the ionization of the nebula is sustained by the quasar radiation 
field.  As described in detail in previous studies (e.g., 
\cite{BBDT83}; \cite{NHJC86}; \cite{MCGPA90}), the power radiated by 
the quasar can easily account for the high ionization level of the 
nebula.  However, the relative symmetry of the envelope requires that 
the ionizing radiation is escaping the quasar symmetrically with 
respect to our line of sight.  The large-scale radiation field from 
\mr\ therefore shows no sign of anisotropy or alignment with the radio 
axis, contrary to expectations from unified models of active galactic 
nuclei that rely on orientation effects (e.g., \cite{B89}; \cite{A93}; 
\cite{UP95}).  This symmetry of the radiation field, if typical of all 
quasars, may have important consequences on our understanding of the 
impact that quasars have on the intergalactic environment (e.g., 
proximity effect of Ly$\alpha$ clouds; \cite{CWSBT82}; \cite{MHPD86}).

\section{Summary and Future Directions}
\label{sec:summary}
Our deep \ha\ observations of \mr\ reveal a spiral complex of ionized
material that extends more or less symmetrically out to $\sim$ 120 kpc
from the quasar.  The coarse velocity field derived from our data
shows a NE -- SW velocity gradient that is opposite to that of the
inner line-emitting region.  The morphology and kinematics of the
nebula cannot be easily explained by an interaction/merger event or a
cooling flow.  We favor the scenario in which the ionized material is
part of a larger neutral envelope that is photoionized by the
radiation field of the quasar.  Deep \hi\ 21-cm observations are
needed to confirm this model.

We suspect that significant quantities of ionized gas may be present
around luminous quasars, but have remained undetected with standard
narrowband imaging techniques.  The observations presented herein
constitute only 40 minutes of integration, yet reach more than an
order of magnitude deeper in flux than previous narrowband imagery.
The TTF instrument allows the observer to tune the filter to a very
precise bandpass in order to match the expected emission and avoid
bright sky features.  The TTF therefore promises to greatly impact a
number of observational programs, such as those that aim at
parameterizing ionized quasar envelopes and quantifying the impact of
quasars on the intergalactic environment.

\acknowledgments 
SV is grateful for support of this research by NSF/CAREER grant
AST-9874973 and by a Cottrell Scholarship awarded by the Research
Corporation.

%
%

%
%
\newpage

\figcaption[fig1.eps]{Deep \ha\ images of the field surrounding the
quasar \mr.  Panels $a$ and $b$ are 1200~second exposures at redshifts
of 0.0640 and 0.0645, respectively, panel $c$ is an I-band continuum
image of the same field, and panel $d$ is a summed \ha\ image. A
bright star (S), a nearby cluster galaxy (G1), and a number of
emission-line knots from Macchetto \etal\ (1990) have been
labeled. All panels are 3\arcmin$\times$2.5\arcmin\ in size, with
north up and east to the left. The lowest contour in panels $a$ and
$b$ represents a surface brightness level of $1.2 \times 10^{-17}$
erg~s$^{-1}$~cm$^{-2}$~arcsec$^{-2}$, while that of panel $d$
corresponds to $1.8 \times 10^{-17}$
erg~s$^{-1}$~cm$^{-2}$~arcsec$^{-2}$.
\label{fig:ttf_ha}}

%
%



\newpage

\begin{thebibliography}{}
\bibitem[Alighieri \etal\ 1984]{APM84} Alighieri, S. D. S., Perryman, 
  M. A. C., \& Macchetto, F. 1984, \apj, 285, 567

\bibitem[Antonucci 1993]{A93} Antonucci, R. 1993, \araa, 31, 473

\bibitem[Bahcall \etal\ 1997]{BKSS97} Bahcall, J. N., Kirhakos, S., 
  Saxe, D. H., \& Schneider, D. P. 1997, \apj, 479, 642

\bibitem[Barthel 1989]{B89} Barthel, P. D. 1989, \apj, 336, 606

\bibitem[Bergeron \etal\ 1983]{BBDT83} Bergeron, J., Boksenberg, A., 
  Dennefeld, M., \& Tarenghi, M. 1983, \mnras, 202, 125

\bibitem[Bland-Hawthorn \& Jones 1998]{BJ98} Bland-Hawthorn, J. \& 
  Jones, D. H. 1998, \pasa, 15, 44

\bibitem[Boisson \etal\ 1994]{BDBP94} Boisson, C., Durret, F., 
  Bergeron, J., \& Petitjean, P. 1994, \aap, 285, 377

\bibitem[Boroson \etal\ 1985]{BPO85} Boroson, T. A., Persson, S. E., 
  \& Oke, J. B. 1985, \apj, 293, 120

\bibitem[Canizares \etal\ 1978]{CMR78} Canizares, C. R., McClintock,
  J. E., \& Ricker, G. R. 1978, \apj, 226, L1

\bibitem[Carswell \etal\ 1982]{CWSBT82} Carswell, R. F., Whelan, J. A. J., 
  Smith, M. G, Boksenberg, A., \& Tytler, D. 1982, \mnras, 198, 91

\bibitem[Cooke \etal\ 1978]{CRMPEWGPMMSPT78} Cooke, B. A., Rickets, M. J.,
  Maccacaro, T., Pye, J. P., Elvis, M., Watson, M. G., Griffiths, R. E.,
  Pounds, K. A., McHardy, I., Maccagni, D., Seward, F. D., Page, C. G.,
  \& Turner, M. J. L. 1978, \mnras, 182, 489

\bibitem[Disney \etal\ 1995]{DBBBCDMMSP95} Disney, M. J., Boyce,
  P. J., Blades, J. C., Boksenberg, A., Cane P., Deharveng, J. M.,
  Macchetto, F., Mackay, C. D., Sparks, W. B., \& Phillipps, S. 1995,
  \nat, 376, 150

\bibitem[Durret \etal\ 1994]{DPPB94} Durret, F., P\'{e}contal, E.,
  Petitjean, P., \& Bergeron, J. 1994, \aap, 291, 392

\bibitem[Fisher \etal\ 1996]{FBKS96} Fisher, K. B., Bahcall, J. N., 
  Kirhakos, S., \& Schneider, D. P. 1996, \apj, 468, 469

\bibitem[Hansen \etal\ 1984]{HNJ84} Hansen, L., N\o rgaard-Nielsen, 
  H. U., \& J\o rgesen, H. E. 1984, \aap, 136, L11

\bibitem[Hutchings \& Campbell 1983]{HC83} Hutchings, J. B. \&
  Campbell, B. 1983, \nat, 303, 584

\bibitem[Hutchings \etal\ 1984a]{HCC84} Hutchings, J. B., Crampton, D., 
  \& Campbell, B. 1984, \apj, 280, 41

\bibitem[Hutchings \etal\ 1984b]{HCCDG84} Hutchings, J. B., Crampton, D., 
  Campbell, B., Duncan, D., \& Glendenning, B. 1984, \apjs, 55, 319

\bibitem[Hutchings \etal\ 1999]{HCMDS99} Hutchings, J. B., Crampton,
  D., Morris, S. L., Durand, D., \& Steinbring, E. 1999, \aj, 117, 1109

\bibitem[Lim \& Ho 1999]{LH99} Lim, J. \& Ho, P. T. P. 1999, \apj, 
  510, L7

\bibitem[Macchetto \etal\ 1990]{MCGPA90} Macchetto, F., Colina, L., 
  Golombek, D., Perryman, M. A. C., \& Alighieri, S. D. S. 1990, 
  \apj, 356, 389

\bibitem[McLure \etal\ 1999]{MDKBOH99} McLure, R. J., Dunlop, J. S.,
  Kukula, M. J., Baum, S. A., O'Dea, C. P., \& Hughes, D. H. 1999,
  preprint astro-ph \#9809030

\bibitem[Miller \etal\ 1996]{MTS96} Miller, J., Tran, H., \& Sheinis, A. 
  1996, \baas, 189, 2004

\bibitem[Mulder \& Valentijn 1992]{MV92} Mulder, P. S. \& Valentijn, 
  E. A. 1992, \aap, 259, 435

\bibitem[Murdoch \etal\ 1986]{MHPD86} Murdoch, H. S., Hunstead, R. W.,
  Pettini, M., \& Blades, J. C. 1986, \apj, 309, 19

\bibitem[N\o rgaard-Nielsen \etal\ 1986]{NHJC86} N\o rgaard-Nielsen, 
  H. U., Hansen, L., J\o rgesen, H. E., \& Christensen, P. R. 1986, 
  \aap, 169, 49

\bibitem[Osterbrock 1989]{O89} Osterbrock, D. E. 1989, Astrophysics of
  Gaseous Nebulae and Active Galactic Nuclei (University Science
  Books: Mill Valley, California)

\bibitem[Phillips 1980]{P80} Phillips, M. M. 1980, \apjl, 236, L45

\bibitem[Ricker \etal\ 1978]{RCDDJDMH78} Ricker, G. R., Clarke, G. W.,
  Doxsey, R. E., Dower, R. G., Jernigan, J. G., Delavaille, J. P.,
  MacAlpine, G. M., \& Hjellming, R. M. 1978, \nat, 271, 35

\bibitem[Stockton \& MacKenty 1987]{SM87} Stockton, A. \& MacKenty,
  J. W. 1987, \apj, 316, 584

\bibitem[Stone \& Baldwin 1983]{SB83} Stone, R. P. S. \& Baldwin,
  J. A. 1983, \mnras, 204, 347

\bibitem[Urry \& Padovani 1995]{UP95} Urry, C. M., \& Padovani,
  P. 1995, \pasp, 107, 803

\bibitem[V\'eron-Cetty \& Woltjer 1990]{VW90} V\'eron-Cetty, M.-P. \&
  Woltjer, L. 1990, \aap, 236, 69

\end{thebibliography}
\end{document}